\documentclass[a4paper,11pt]{article}
\pdfoutput=1 

\usepackage{jheppub} 
\usepackage[T1]{fontenc} 
\usepackage{mathrsfs}
\usepackage{graphicx}
\usepackage{subcaption}
\usepackage{tikz}
\usepackage{tabularx}
\usepackage{array}
\usepackage{longtable}
\usepackage{booktabs}
\usepackage{graphicx,tikz} 
\title{\boldmath Large-$N$ Dynamics of a QCD-Inspired Unitary Matrix Model}

\author[a]{Anuj Malik}

\affiliation[a]{Department of Physics, Malaviya National Institute of Technology, \\Jaipur-302017, INDIA}

\emailAdd{2021rpy9096@mnit.ac.in}

 \abstract{We study the large-$N$ limit of $U(N)$ and $SU(N)$ unitary matrix models inspired by QCD. The model is analyzed in two cases: $\mu = 0$, where the potential is real, and finite $\mu$, where it becomes complex. The complex action drives the eigenvalues into the complex plane, leading to $\langle U \rangle \neq \langle U^{-1} \rangle$.
In the ungapped phase, we obtain analytic expressions for the spectral density, Wilson loops, and free energy, which reproduce the low-temperature behaviour of QCD. In contrast, the gapped phase involves a nontrivial resolvent and is solved partially analytically and numerically. At $\mu=0$, the model exhibits a $3^{rd}$ order phase transition, while at finite $\mu$, it shows a continuous phase transition of at least second order.}

\begin{document} 
\maketitle
\flushbottom

\section{Introduction}\label{sec:Intro}
The study of random matrix ensembles plays a key role in theoretical physics and applied mathematics. They appear in many areas, including quantum field theory, condensed matter physics, and statistics \cite{Mehta1991, log-gases_2010, Baik2016, guhr_random-matrix_1998}. They naturally arise in gauge theories \cite{gross_possible_1980, aharony2004, dumitru2005, hands_qcd_2010}, two-dimensional gravity and string theory \cite{brezin_exactly_1990, douglas_strings_1990, gross_nonperturbative_1990}, and statistical systems such as the Ising model and disordered conductors \cite{mccoy_two-dimensional_1973, muttalib_new_1993, deift2013}. These models are also related to integrable structures such as Painlevé equations and show rich phase behavior that has been widely studied \cite{mandal_phase_1990, russo_phases_2020, santilli_exact_2020, Russo2020MultiplePI, santilli_multiple_2022, rossi_large-n_1998, marino_houches_2005}.

A key advantage of matrix models is that they often allow a fully non-perturbative analysis. This makes them useful for studying resurgent and trans-series structures \cite{marino2008, ahmed_transmutation_2017, Anees2018}. They have also been used to study the \emph{sign problem}, which makes standard Monte Carlo simulations difficult in lattice gauge theory \cite{hands_simulating_2007}. Methods such as complex Langevin dynamics \cite{aarts2010, boguslavski2025, klauder_coherent-state_1984} and Lefschetz thimble approaches \cite{fujii_monte_2015, cristoforetti2013, nishimura2024} have been proposed to address this issue, and matrix models provide simple settings in which these methods can be tested.

The partition function of a U($N$) matrix model is
\begin{equation} \label{def:partitionFunction}
    Z = \int_{\mathrm{U}(N)} dU \, e^{-\mathcal{S}(U)},
\end{equation}
where $dU$ denotes Haar measure on U($N$) normalized to one, and $\mathcal{S}$ is the action. Observable expectation values are computed as
\begin{equation} \label{def:expectationValue}
    \langle f(U) \rangle = \frac{1}{Z} \int_{\mathrm{U}(N)} dU \, f(U)\, e^{-\mathcal{S}(U)}.
\end{equation}
In this work, we consider actions of the form $\mathcal{S}=N\,\mathrm{Tr}\,V(U)$ with the potential
\begin{equation} \label{def:potential}
    V(z) = -a\,z-b\,z^{-1} - \sigma \left[\ln(1+\alpha z)+\ln\!\left(1+\frac{1}{\gamma z}\right)\right].
\end{equation}

For real potentials, the symmetry of the Haar measure implies $\langle U \rangle = \langle U^{-1} \rangle$. However, in our case, the potential is complex, so this relation does not generally hold~\cite{Basu2018}. The logarithmic terms introduce Fisher--Hartwig-type singularities~\cite{fahs_uniform_2021}, which lead to a nontrivial phase structure.

A useful interpretation of the potential~\eqref{def:potential} is as a low-temperature effective potential arising from the one-loop effective description of QCD on a sphere, originally obtained from $SU(N)$ gauge theories \cite{Aharony:2003sx, hands_qcd_2010}. The linear terms in $V(z)$ arise from the bosonic sector, while the logarithmic terms originate from the fermionic contribution. The parameters $\alpha$ and $\gamma$ correspond to the fugacities of quarks and antiquarks, respectively. The parameters $a,\,b,\,\sigma$ play the role of ’t Hooft-like couplings, as discussed in Sec.~\ref{sec:Truncation} and the references therein.

This model reduces to several well-known cases in appropriate limits. For $\sigma = 0$, it reduces to the $\mathbf{ab}$-model of \cite{Basu2018}, and for $a = b$ further to the Gross--Witten--Wadia model \cite{gross_possible_1980}.
For $a = b = 0$ and $\gamma \to \infty$, it reproduces the model of Ref.~\cite{hands_qcd_2010}, describing a two-level QCD system on a compact space; and for $\alpha = 1/\gamma$, it reduces to the model studied in \cite{santilli_exact_2020}, while for $a = b$ and $\alpha = \gamma = 1$, it reproduces the model of \cite{russo_multiple_2020}.

Finally, for $a = b$ and $\alpha = 1/\gamma$, the model reduces to the $\mu = 0$ limit of QCD. Therefore, our model connects several known models within a single framework.
\\
\\
In the next section, we present the formulation of the effective potential \eqref{def:potential}.

\section{Background of the Potential}\label{sec:Truncation}

In this section, we review the structure of the potential \eqref{def:potential} and its relation to the one-loop effective description of QCD on a small sphere ($S^{1} \times S^{3}$). This formulation was originally obtained for $SU(N)$ gauge theories with more general matter content in Refs.~\cite{Aharony:2003sx,hands_qcd_2010}. We begin with the one-loop effective action, expressed in terms of the eigenvalues ${e^{i\theta_i}}$ of the Polyakov loop. The partition function then takes the form
\begin{equation}
Z_{\mathrm{QCD}} = \int [d\theta] \; e^{-S(\theta_i)},
\qquad
S(\theta_i) = S_b(\theta_i) + S_f(\theta_i),
\end{equation}
where $[d\theta]=\prod_{i=1}^{N}d\theta_{i}$ and the bosonic and fermionic contributions are
\begin{align}
S_b(\theta_i) &= \sum_{i,j=1}^{N} \sum_{n=1}^{\infty}
\frac{1}{n}\left(1 - z_b(n\beta/R)\right)
\cos\!\big(n(\theta_i - \theta_j)\big), \\
S_f(\theta_i) &= \sum_{n=1}^{\infty} \frac{(-1)^n}{n} N_f \,
z_f(n\beta/R,mR) \sum_{i=1}^{N} \left[e^{n\beta\mu + i n\theta_i} + e^{-n\beta\mu - i n\theta_i}\right].
\end{align}
Here $z_b(n\beta/R)$ and $z_f(n\beta/R,mR)$ are the single particle partition functions for bosons and fermions
\begin{equation}
z_b(n\beta/R)=2\sum_{l=1}^{\infty} l(l+2)e^{-n\beta(1+l)/R},
\qquad
z_f(n\beta/R,mR)=2\sum_{l=1}^{\infty} l(1+l)e^{-n\beta \varepsilon_l},
\end{equation}
with
\begin{equation}
\varepsilon_l = \sqrt{m^2 + \left(l+\tfrac12\right)^2 R^{-2}}.
\end{equation}
To rewrite the theory in a compact matrix form, we introduce the Polyakov loop matrix $U = \mathrm{diag}(e^{i\theta_1}, \dots, e^{i\theta_N})$, so that
\begin{equation}
\sum_{i,j=1}^N \cos\big(n(\theta_i - \theta_j)\big)
= \mathrm{Tr}\, U^n \, \mathrm{Tr}\, U^{\dagger n}.
\end{equation}
The unimportant constant term in $S_b(U)$ is proportional to the Vandermonde determinant and is absorbed into the Haar measure. The remaining bosonic contribution then becomes
\begin{equation}
S_b(U) = -\sum_{n=1}^{\infty} \frac{z_b(n\beta/R)}{n} \,
\mathrm{Tr}\, U^n \, \mathrm{Tr}\, U^{\dagger n}.
\end{equation}
The fermionic contribution can be written as
\begin{equation}
S_f(U) =-\sum_{l=1}^{\infty} \xi_l\left[\mathrm{Tr} \ln\big(1 + \alpha_l U\big)+ \mathrm{Tr} \ln\big(1 +  \gamma_{l}^{-1} U^\dagger\big)\right],
\end{equation}
where
\begin{equation}
\xi_l = 2l(l+1)N_f,
\qquad
\alpha_l = e^{\beta(\mu - \varepsilon_l)},
\qquad
\gamma_{l} = e^{\beta(\mu + \varepsilon_l)}.
\end{equation}
Combining the bosonic and fermionic contributions, the effective action takes the form
\begin{equation}\label{eq:original_potential}
S(U) =-\sum_{n=1}^{\infty} \frac{z_b(n\beta/R)}{n} \,\mathrm{Tr}\, U^n \, \mathrm{Tr}\, U^{\dagger n}-\sum_{l=1}^{\infty} \xi_l\left[\mathrm{Tr} \ln\big(1 + \alpha_l U\big)+ \mathrm{Tr} \ln\big(1 + \gamma_{l}^{-1} U^\dagger\big)\right].
\end{equation}
Although the action \eqref{eq:original_potential} is symmetric under $U \to U^\dagger$ for $\xi_l = 0$ and $\mu = 0$, it is difficult to solve analytically in the large-$N$ limit. Therefore, we introduce controlled approximations.
\paragraph{(i) Leading winding truncation:} In the low-temperature regime, the $n=1$ winding mode dominates, and the bosonic contribution simplifies to
\begin{equation}
S_b(U) \approx -z_b(\beta/R)\, \mathrm{Tr}\, U \, \mathrm{Tr}\, U^\dagger.
\end{equation}
The double-trace term is approximated by a single-trace contribution using $\mathrm{Tr}\, U \, \mathrm{Tr}\, U^\dagger\approx\langle \mathrm{Tr}\, U \rangle \mathrm{Tr}\, U^\dagger+\langle \mathrm{Tr}\, U^\dagger \rangle \mathrm{Tr}\, U-\langle \mathrm{Tr}\, U \rangle \langle \mathrm{Tr}\, U^\dagger \rangle$. For further details, see Ref.~\cite{Hollowood:2012nr}, where this approximation is discussed.
\\
\\
\textbf{(ii) Lowest fermion mode truncation ($l=1$).} 
\\
\\
We now introduce the 't Hooft-like coupling parameters of the truncated action as
\begin{equation}
a = \frac{z_b(\beta/R)}{N} \langle \mathrm{Tr}\, U^\dagger \rangle,
\quad
b = \frac{z_b(\beta/R)}{N} \langle \mathrm{Tr}\, U \rangle,
\quad
\sigma = \frac{\xi_1}{N},
\end{equation}
and
\begin{equation}
\alpha = e^{\beta(\mu - \varepsilon)},
\qquad
\frac{1}{\gamma} = e^{-\beta(\mu + \varepsilon)},
\end{equation}
With these definitions, the action \eqref{eq:original_potential} reduces to the truncated form given by
\begin{equation}
S(U) =- a\, \mathrm{Tr}\, U- b\, \mathrm{Tr}\, U^\dagger- \sigma \, \mathrm{Tr} \ln\big(1 + \alpha U\big)- \sigma \, \mathrm{Tr} \ln\!\left(1 + \frac{1}{\gamma} U^\dagger\right).
\end{equation}

In the next section, we present the theoretical framework for analyzing the large-$N$ limit of the unitary matrix model.

\section{Theoretical Framework}
In this section, we summarize the standard approach to solving U($N$) and SU($N$) matrix models, following Ref.~\cite{hands_qcd_2010}, and repeat the main steps for the reader’s convenience. The partition function \eqref{def:partitionFunction} is defined as an integral over U($N$), where SU($N$) is imposed by the constraint $\det U = 1$, or equivalently $\mathrm{Tr}\,\ln U = 0$.

It is convenient to replace the constrained SU($N$) with an unconstrained U($N$) integral, with the $\det U = 1$ condition enforced by a Lagrange multiplier $\mathscr{N}$. In the large-$N$ limit, this induces the shift
\begin{equation}
V(z) \;\rightarrow\; V(z) + \mathscr{N}\ln z, \qquad \mathscr{N}\in\mathbb{R}.
\end{equation}

Upon diagonalizing $U$, the eigenvalues are written as $e^{i\theta_i}$ with $\theta_i\in[-\pi,\pi)$, $i=1,\dots,N$. The U$(N)$ integral then reduces to
\begin{equation}
Z = \int \prod_{i=1}^{N} d\theta_i \, e^{-\mathcal{S}_{\mathrm{eff}}},
\end{equation}
with effective action
\begin{equation}\label{effectiveAction}
\mathcal{S}_{\mathrm{eff}}
= N \sum_{i=1}^{N} V(e^{i\theta_i})
- \frac{1}{2}\sum_{i,j=1}^{N}\ln\!\left[\sin^2\!\left(\frac{\theta_i-\theta_j}{2}\right)\right]
+ iN\mathscr{N}\sum_{i=1}^{N}\theta_i .
\end{equation}
The second term represents the Vandermonde interaction among eigenvalues. In this form, the integral can be evaluated by the saddle-point method, which becomes exact in the strict large-$N$ limit. The saddle-point equations, obtained from $\partial \mathcal{S}_{\mathrm{eff}}/\partial \theta_i=0$, read
\begin{equation}
V'(\theta_i) + i\mathscr{N}
= \frac{1}{N}\sum_{j\neq i}\cot\!\left(\frac{\theta_i-\theta_j}{2}\right).
\end{equation}
Since the left-hand side is generically complex, the saddle-point configurations need not lie on the real $\theta$ axis.

We now introduce the spectral density
\begin{equation}
\bar{\rho}(\theta)=\frac{1}{N}\sum_{i}\delta(\theta-\theta_i),
\end{equation}
which, in the large-$N$ limit, allows the replacement
\begin{equation}
\frac{1}{N}\sum_{i=1}^{N} f(\theta_i)\;\longrightarrow\;
\int d\theta\, \bar{\rho}(\theta)\, f(\theta).
\end{equation}
By construction, $\bar{\rho}(\theta)$ is normalized to unity. Introducing $z_i=e^{i\theta_i}$, the saddle-point equation becomes
\begin{equation}
z_i V'(z_i)+\mathscr{N}
= \frac{1}{N}\sum_{j\neq i}\frac{z_i+z_j}{z_i-z_j}.
\end{equation}
In the continuum large-$N$ limit, this turns into
\begin{equation}\label{EOM}
zV'(z)+\mathscr{N}
= \mathscr{P}\!\int_{\mathcal C}\frac{dz'}{2\pi i z'}\,\rho(z')\,\frac{z+z'}{z-z'},
\qquad z\in\mathcal C ,
\end{equation}
where $\mathscr{P}$ denotes the Cauchy principal value and $\rho(z):=2\pi \bar{\rho}(\theta)$ for $z\in\mathcal C$. The contour $\mathcal {C} $ is the support of the eigenvalue distribution. The corresponding density along $\mathcal C$ is
\begin{equation}
D(z)=\frac{\rho(z)}{2\pi i z}, \qquad z\in\mathcal C .
\end{equation}
Normalization follows from that of $\bar{\rho}(\theta)$,
\begin{equation}\label{spectralFunctionNormalization}
\int_{\mathcal C}\frac{dz}{2\pi i z}\,\rho(z)=1.
\end{equation}
We extend $\rho(z)$ to the complex plane by taking the equation of motion as its defining relation. The problem is then to determine the spectral density $\rho(z)$ and its support $\mathcal{C}$ for a given potential. This defines a Riemann--Hilbert problem \cite{brezin1978, marino_houches_2005}, which can be solved explicitly for single-cut configurations where $\mathcal{C}$ is connected.

For a real potential in a U$(N)$ model ($\mathscr{N}=0$), the saddle points lie on the unit circle and $\mathcal{C}$ is either the full circle or an arc thereof. In the SU$(N)$ model, however, the eigenvalues are not constrained to remain on the unit circle, and the support may extend into the complex plane.

The contour $\mathcal{C}$ follows from the differential equation
\begin{equation}\label{Contour}
\frac{dz}{d\theta}=\frac{i z}{\rho(z)}, \qquad z \equiv z(\theta).
\end{equation}
In the large-$N$ limit, the expectation value \eqref{def:expectationValue} becomes
\begin{equation}
\langle f(U) \rangle
= \int_{\mathcal C}\frac{dz}{2\pi i z}\,\rho(z)\,f(z).
\end{equation}
Finally, the spectral density must satisfy the SU$(N)$ constraint. Using $\det U=1$, one finds
\begin{equation}\label{eq:SU}
\det U = 1
\;\Rightarrow\;
\sum_i \theta_i = 0
\;\Rightarrow\;
\int_{\mathcal C}\frac{dz}{2\pi i z}\,\rho(z)\,\ln z = 0 .
\end{equation}
\\
We now define the \textbf{ungapped phase} as the regime in which the spectral density has compact, closed support $\mathcal{C}$. In this case, the eigenvalue distribution forms a closed contour, and the saddle-point equation \eqref{EOM} reduces to
\begin{equation}\label{eq:EOM_ungapped}
zV'(z)+\mathscr{N}
= \mathscr{P}\!\oint_{\mathcal C}\frac{dz'}{2\pi i z'}\,\rho(z')\,\frac{z+z'}{z-z'},
\qquad z\in\mathcal C .
\end{equation}

Guided by the analytic structure of $zV'(z)$, we assume the following ansatz for the spectral density:
\begin{equation}
\rho(z)
= \sum_{n=0}^{\infty} \frac{A_n}{z^n}
+ \sum_{\text{poles } z_p}\sum_{n=1}^{\infty} \frac{B_{p,n}}{(z - z_p)^n}.
\end{equation}

In particular, it is necessary to include a pole at $z=0$ in order to satisfy the normalization condition \eqref{spectralFunctionNormalization}. This implies that $z=0$ must lie inside the contour $\mathcal{C}$. Other poles may lie either inside or outside the contour, depending on their magnitudes.

The coefficients $A_n$ and $B_{p,n}$ are determined by substituting the ansatz into \eqref{eq:EOM_ungapped} and matching both sides order by order. This matching procedure also fixes the Lagrange multiplier $\mathscr{N}$. In addition, the normalization condition can be used to determine one remaining coefficient.

Once the spectral density is determined, all observables in this phase can be computed from it. The ungapped phase is physically valid only if $\rho(z) > 0$ on $\mathcal{C}$. As parameters are varied, $\rho(z)$ may vanish at a point on the contour, typically at $z(\pm\pi)$ on the negative real axis, signaling the onset of a gap. At this point, the support $\mathcal{C}$ splits, and the system transitions into the \textbf{gapped phase}, similar to the GWW model \cite{gross_possible_1980}.

The gapped phase can be most effectively investigated in terms of the resolvent,
\begin{equation}
    R(z) = \Big\langle \frac{z+U}{z-U} \Big\rangle .
\end{equation}
This definition implies the asymptotic limits
\begin{equation}\label{endpoint}
    R(z) =
    \begin{cases}
        -1, & z \to 0, \\
        +1, & z \to \infty .
    \end{cases}
\end{equation}
At large-$N$, the resolvent admits the integral representation
\begin{equation}\label{Resolvent}
    R(z) = \int_{\mathcal{C}} \frac{dz'}{2\pi i z'} \, \rho(z') \, \frac{z+z'}{z-z'} .
\end{equation}
It follows that $R(z)$ is analytic in the complex plane except along the support $\mathcal{C}$, where it develops a branch cut with endpoints $\mathfrak{z}$ and $\mathfrak{z}^*$.

Using the Sokhotski--Plemelj formula \cite{muskhelishvili2013}, one obtains the relations
\begin{equation}\label{CutEOM}
    z V'(z) = \frac{1}{2} \lim_{\epsilon \to 0} \big[ R(z+\epsilon) + R(z-\epsilon) \big],
\end{equation}
and
\begin{equation}\label{eq:densityFromResolvent}
    \rho(z) = \frac{1}{2} \, \Delta_{\mathcal{C}} R(z),
\end{equation}
where $\Delta_{\mathcal{C}} R(z)$ denotes the discontinuity of the resolvent across $\mathcal{C}$.

Motivated by these relations and by the fact that $\mathcal{C}$ serves as a branch cut for $R(z)$, we adopt the ansatz
\begin{equation}
    R(z) = z V'(z) + g(z)\, Q(z),
\end{equation}
where $V(z)$ is the modified potential including the Lagrange multiplier. The function $g(z)$ is analytic on $\mathcal{C}$ and may be meromorphic elsewhere, while
\begin{equation}
    Q(z) = \sqrt{z-\mathfrak{z}}\, \sqrt{z-\mathfrak{z}^*},
\end{equation}
encodes the discontinuity across $\mathcal{C}$, with the branch cut chosen along $\mathcal{C}$. This choice ensures $\rho(\mathfrak{z}) = \rho(\mathfrak{z}^*) = 0$. Substituting into \eqref{eq:densityFromResolvent}, we obtain
\begin{equation}
    \rho(z) = \frac{1}{2}\, g(z)\, \Delta_{\mathcal{C}} Q(z).
\end{equation}
Hence, determining the spectral density reduces to solving for $g(z)$, and the endpoints $\mathfrak{z}$ and $\mathfrak{z}^*$. To determine the function $g(z)$, we first note that \eqref{eq:densityFromResolvent} allows the expectation value of any function $f(U)$ to be written as a closed contour integral,
\begin{equation}\label{eq:openToClosed}
    \langle f(U) \rangle
    = \int_{\mathcal{C}} \frac{dz}{2\pi i z} \, \rho(z)\, f(z)
    = \oint_{\mathcal{C}^*} \frac{dz}{4\pi i z} \, R(z)\, f(z),
\end{equation}

\begin{figure}[t]
    \centering
    \includegraphics[width=.47\textwidth]{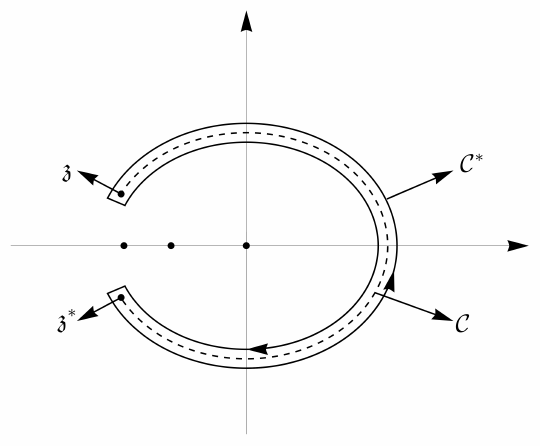}
    \qquad
    \includegraphics[width=.47\textwidth]{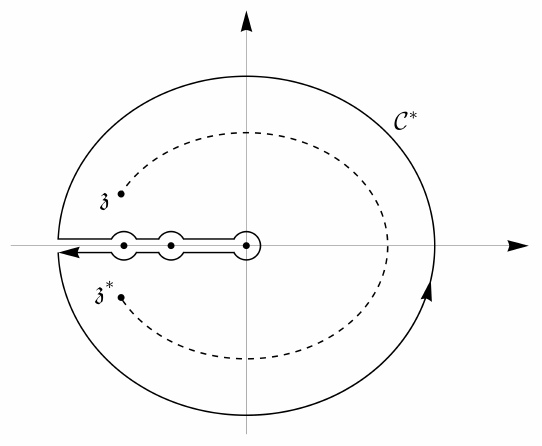}
    \caption{\small Schematic of the support $\mathcal C$ in the gapped phase (dashed). The dots on the negative real axis indicate the potential's poles. The solid curve denotes the auxiliary contour $\mathcal C^\ast$ used to compute integrals in the gapped phase, which can be continuously deformed to the contour on the right.}
    \label{fig:deformed contour}
\end{figure}

where $\mathcal{C}^*$ denotes a closed contour encircling the support $\mathcal{C}$, as shown in the left panel of Fig.~\ref{fig:deformed contour}. Using this representation, one obtains the general expression 
\begin{equation}\label{eq:g(z)}
    g(z)
    = \oint_{\mathcal{C}^*} \frac{dz'}{2\pi i} \,
    \frac{z' V'(z')}{(z' - z)\, Q(z')} .
\end{equation}
This result is closely related to a theorem due to Tricomi \cite{tricomi1985}. The contour $\mathcal{C}^*$ can be continuously deformed to the one shown in the right panel of Fig.~\ref{fig:deformed contour}. As a consequence, the integral in \eqref{eq:g(z)} receives contributions only from the residues at the poles and at infinity. Evaluating these residues yields $g(z)$. Once the spectral density is determined, all observables in this phase can be computed from it.

\section{Effective model at $\mu = 0$}

The effective potential of the model is given by
\begin{equation}\label{def:mu=0_potential}
V(z) = - a\, (z + z^{-1})- \sigma \left[ \ln\big(1 + \alpha z\big)+ \ln\big(1 + \alpha z^{-1}\big) \right].
\end{equation}
Here, the parameters $a$ and $\sigma$ are taken to be real and positive, while $\alpha$ satisfies $0 < \alpha \leq 1$. As the potential is real, we restrict our analysis to the $U(N)$ model, for which the Lagrange multiplier $\mathscr{N}$ vanishes.

\subsection{Ungapped Phase}
Solving the saddle-point equation \eqref{eq:EOM_ungapped}, one obtains the spectral density 
\begin{equation}
    \rho(z)=1+a(z+z^{-1})+\sigma\left(1+\dfrac{\alpha}{\alpha+z}-\dfrac{1}{1+\alpha\,z}\right)
\end{equation}
The support $\mathcal{C}$ of the eigenvalue distribution is obtained using Eq. \eqref{Contour}
\begin{equation}\label{eq:Contour_at_mu_zero}
     |z|^{1+\frac{1}{\sigma}}\,e^{\frac{a}{\sigma}\, \cos{\theta} \left( |z| - \frac{1}{|z|} \right)} = |\alpha\,z + 1|^{-1} \,|z +\alpha|
\end{equation}
Since $z = |z| e^{i\theta}$, Eq.~\eqref{eq:Contour_at_mu_zero} is satisfied for all $\theta$ only when $|z| - |z|^{-1} = 0$, implying $|z|=\pm1$; hence, the support $\mathcal{C}$ is the unit circle.

 As $a$ is varied, the spectral density may develop a zero on the negative real axis at $z=-1$, signaling the onset of a gap. At this point, the support $\mathcal{C}$ splits, and the system undergoes a transition to the gapped phase. Imposing the condition $\rho(-1)=0$ determines the phase boundary,
\begin{equation}\label{eq:Phase_at_zero_chemical}
    a = \dfrac{1}{2}-\dfrac{\sigma \alpha}{1-\alpha}.
\end{equation}
The winding Wilson loops are
\begin{equation}\label{eq:Wilson_at_zero_chemical}
W_n := \frac{1}{N}\langle \mathrm{tr}\,U^n\rangle
= \oint_{\mathcal C}\frac{dz}{2\pi i z}\,\rho(z)\,z^n
= 
\begin{cases}
a+\sigma\,\alpha, & n= \pm\,1,\\[6pt]
(-1)^{n+1}\sigma\,\alpha^n, & n> 1, \\[6pt]
(-1)^{|n|+1}\sigma\,\alpha^{|n|}, & n< -1 .
\end{cases}
\end{equation}
We now consider the derivatives of the free energy,
\begin{equation}
\frac{\partial F}{\partial t}
=\Big\langle \frac{\partial V(U)}{\partial t}\Big\rangle 
= \oint_{\mathcal C}\frac{dz}{2\pi i z}\,\rho(z)\,
\frac{\partial V(z)}{\partial t},
\qquad 
t= a,\sigma,\alpha .
\end{equation}
\begin{align}
 \frac{\partial F}{\partial a} &= -2(a+\sigma\alpha), \qquad
 \frac{\partial F}{\partial \sigma} = -2(a\,\alpha-\sigma\ln{(1-\alpha^2)}), \qquad
 \frac{\partial F}{\partial \alpha}= -2\sigma(a+\dfrac{\alpha\sigma}{1-\alpha^2}) .
\end{align}
Integrating these relations, one finds that the free energy is given by
\begin{equation}
F=-a^2-2a\sigma\alpha+\sigma^2\ln{(1-\alpha^2)} .
\end{equation}
\textbf{Remarks:} As discussed in Sec.~\ref {sec:Intro}, the observables of the model reduce to the expected observables in the relevant limits and show that the model consistently reproduces known cases.

Using the relation $a = z_b(\beta/R)\, W_{-1}$ derived in Sec.~\ref{sec:Truncation}, and substituting it into Eq.~\eqref{eq:Wilson_at_zero_chemical}, we obtain the self-consistent expression for the Wilson loop at $\mu = 0$:
\begin{equation}\label{eq:Wilson_mu_0}
W_{\pm 1} = \frac{\sigma \alpha}{1 - z_b(\beta/R)} \, .
\end{equation}
Similarly, the phase boundary Eq. \eqref{eq:Phase_at_zero_chemical} in terms of $z_b$ is given by
\begin{equation}\label{eq:Phase_mu_0}
z_b = \frac{1 - \alpha - 2 \alpha \sigma}{1 - \alpha - 2 \alpha^2 \sigma} \, .
\end{equation}
In the special case $\sigma = 0$, Eqs.~\eqref{eq:Wilson_mu_0} and \eqref{eq:Phase_mu_0} reduce to $W_{\pm 1} = 0$ and $z_b = 1$ which are consistent with \cite{Damgaard:1986mx, Aharony:2003sx}.

\begin{figure}[t]
    \centering
    \includegraphics[width=0.6\linewidth]{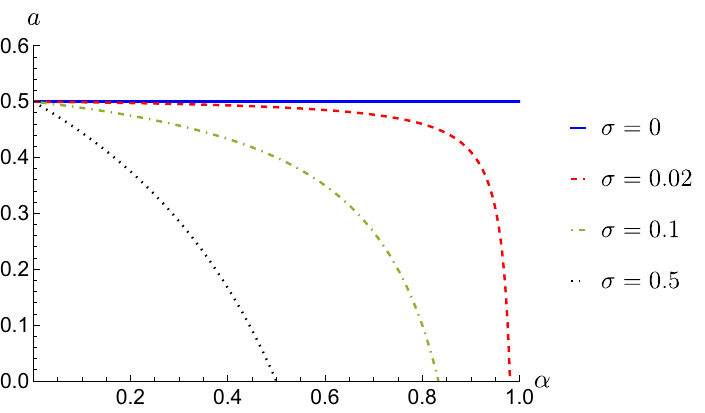}
    \caption{\small Critical parameters governing the ungapped–gapped phase transition. Below these lines, the system is in the ungapped phase, while above them, the system is in the gapped phase.}
    \label{fig:mu_0_phase_diagram}
\end{figure}

\subsection{Gapped Phase}
Using Eq.~\eqref{eq:g(z)}, the function $g(z)$ is obtained as
\begin{equation}
g(z)= a\left(1+\frac{1}{|\mathfrak{z}|\,z}\right)
+\sigma\left(\frac{1}{A(1+\alpha z)}+\frac{\alpha}{B(\alpha+z)}\right).
\end{equation}
Here, $|\mathfrak{z}|=1$ indicates that the endpoint $\mathfrak{z}$ lies on the unit circle. We have also introduced the shorthand notations
\begin{equation}
A \equiv \left|\frac{1}{\alpha}+\mathfrak{z}\right|, 
\qquad 
B \equiv \left|\alpha+\mathfrak{z}\right|.
\end{equation}
The remaining parameters $\mathfrak{z}$ and its complex conjugate $\mathfrak{z}^\ast$ are determined by imposing two independent conditions arising from the asymptotic behaviour of the resolvent, Eq.~\eqref{endpoint}. These constraints take the form
\begin{equation}\label{eq:endpoint Constraints}
\begin{aligned}    
-a\left(1-\mathrm{Re}\,\mathfrak{z}\right)
+\sigma-\sigma \!\left(\frac{1}{A}+\frac{1}{B}\right) &= -1, 
\\[6pt]
a\left(1-\mathrm{Re}\,\mathfrak{z}\right)
-\sigma+\sigma\!\left(\frac{1}{\alpha A}+\frac{\alpha}{B}\right) &= 1 .
\end{aligned}
\end{equation}
Since $A = B/\alpha$, the two constraints in \eqref{eq:endpoint Constraints} are not independent and reduce to a single equation,
\begin{equation}\label{eq:Find_Real_z_mu0}
a\left(1-\mathrm{Re}\,\mathfrak{z}\right)
- \sigma\left(1- \frac{1+\alpha}{B}\right)= 1,
\end{equation}
where
\begin{equation}
B = \sqrt{1+\alpha^2+2\alpha\,\mathrm{Re}\,\mathfrak{z}}.
\end{equation}
Thus, the problem reduces to solving for $\mathrm{Re}\,\mathfrak{z}$. Equation~\eqref{eq:Find_Real_z_mu0} admits a unique solution; however, it cannot be expressed in a simple closed form. The Wilson loop expectation values are obtained using Eq.~\eqref{eq:openToClosed} as
\begin{equation}
W_{\pm1} = \frac{a}{2}\left(1-\mathrm{Re}\,\mathfrak{z}+\frac{(\mathrm{Im}\,\mathfrak{z})^2}{2}\right)+ \frac{\sigma}{2}\left(\frac{1}{\alpha }+\alpha\right) -\frac{\sigma}{2 B} \left(\frac{1}{\alpha} + \mathrm{Re}\, \mathfrak{z} \right)- \frac{\sigma\, \alpha}{2 B} \left(\alpha + \mathrm{Re}\, \mathfrak{z} \right).
\end{equation}
For completeness, we also consider derivatives of the free energy. As an example, the derivatives with respect to $a$ and $\alpha$ are given by
\begin{equation}
\frac{\partial F}{\partial a} = -2W_{\pm 1}.
\end{equation}
\begin{equation}
\begin{split}
      \frac{\partial F}{\partial \alpha}= \dfrac{\sigma\,a}{2}\left(\dfrac{1}{\alpha^2}-2+\text{Re}\,\mathfrak{z}-\dfrac{(1-\alpha)B}{\alpha^2}\right) 
      - \frac{\sigma ^2\alpha}{1+\alpha } -\frac{\sigma^2}{2B}\left(1- \text{Re} \, \mathfrak{z} \right)
\end{split}
\end{equation}
The free energy cannot be expressed in closed form owing to its implicit dependence on $\mathrm{Re}\,\mathfrak{z}$.
\\
\\
\textbf{Remarks:} As discussed in Sec.~\ref{sec:Intro}, the observables of the model reduce to the expected observables in the relevant limits, showing that the model consistently reproduces known cases.

As discussed above, $\mathrm{Re}\,\mathfrak{z}$ is not available in closed form and depends implicitly on $a$. Consequently, it is not possible to derive an explicit self-consistent form for the Wilson loop at $\mu = 0$ in this phase. However, in the limiting case of the GWW model, one finds
\begin{equation}
\mathrm{Re}\,\mathfrak{z} = 1 - \frac{1}{a}.
\end{equation}
Using the constraint $(\mathrm{Re}\,\mathfrak{z})^2 + (\mathrm{Im}\,\mathfrak{z})^2 = 1$, the Wilson loop expectation value is obtained as
\begin{equation}
W_{\pm 1} = 1 - \frac{1}{4a}.
\end{equation}
Furthermore, using the relation $a = z_b(\beta/R)\, W_{\pm1}$, one obtains
\begin{equation}
W_{\pm 1} = \frac{1}{2}\left(1 + \sqrt{1 - \frac{1}{z_b}}\right),
\end{equation}
in agreement with the corresponding results in Yang--Mills theory \cite{Damgaard:1986mx, Aharony:2003sx}.

\subsection{Phase transition}
To determine the order of the phase transition, one must analyze the behavior at the critical point. The transition occurs when
\begin{equation}
\mathrm{Re}\,\mathfrak{z} \to -1,\qquad \mathrm{Im}\,\mathfrak{z} \to 0.
\end{equation}
The Wilson loop and its first derivatives are continuous at the transition point
\begin{equation}
W_{\pm 1}^{\text{Ungap}} = W_{\pm 1}^{\text{Gap}} = a + \sigma\,\alpha,\qquad
\frac{\partial W_{\pm 1}^{\text{Ungap}}}{\partial a}=\frac{\partial W_{\pm 1}^{\text{Gap}}}{\partial a}= 1, 
\qquad
\frac{\partial W_{\pm 1}^{\text{Ungap}}}{\partial \sigma}=\frac{\partial W_{\pm 1}^{\text{Gap}}}{\partial \sigma}= \alpha.
\end{equation}
However, the second derivatives of the Wilson loop vanish in the ungapped phase, while they remain finite in the gapped phase, resulting in a discontinuity at the transition point. This behavior signals a third-order phase transition across all parameters, in close analogy with the GWW model \cite{gross_possible_1980}. Consistently, the third derivative of the free energy is also discontinuous across the transition.

In Figure~\ref{fig:mu_0_phase_diagram}, we observe that there is no phase transition for $a > 1/2$  in the region of positive $\sigma$. Notably, $a = 1/2$ corresponds to the critical value in the GWW model. This indicates that including the logarithmic term in the model lowers the critical threshold relative to the GWW value.

\section{Effective model at finite $\mu$ }

The effective potential of the model is defined as
\begin{equation}\label{def:mu=0 potential}
    V(z) = - a\,z - b\,z^{-1}
    - \sigma\left( \ln\big(1 + \alpha z\big)
    + \ln\!\left(1 + \frac{1}{\gamma z}\right)\right).
\end{equation}
The parameters $a$, $b$, and $\sigma$ are restricted to positive real values. For $\mu < \varepsilon$, the parameter $\alpha$ lies in the interval $0 < \alpha \leq 1$, while for $\mu > \varepsilon$, it lies in the range $1 \leq \alpha < \infty$. The parameter $\gamma$ satisfies $1 \leq \gamma < \infty$. As the potential is complex, the $\mathrm{SU}(N)$ constraint must be imposed.

\subsection{Ungapped Phase}
\subsubsection{Small $\alpha$ ($\mu<\varepsilon$)}
\begin{figure}[t]
    \centering
    \includegraphics[width=0.49\linewidth]{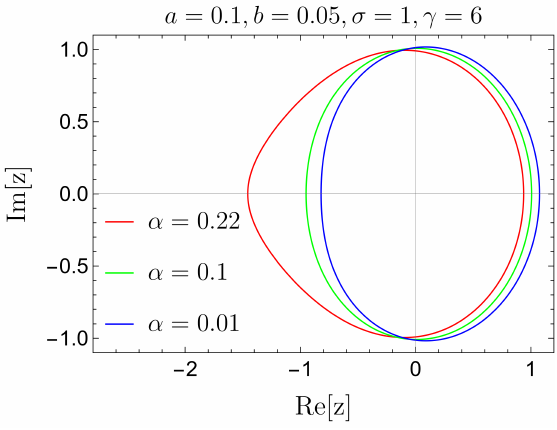}
    \includegraphics[width=0.48\linewidth]{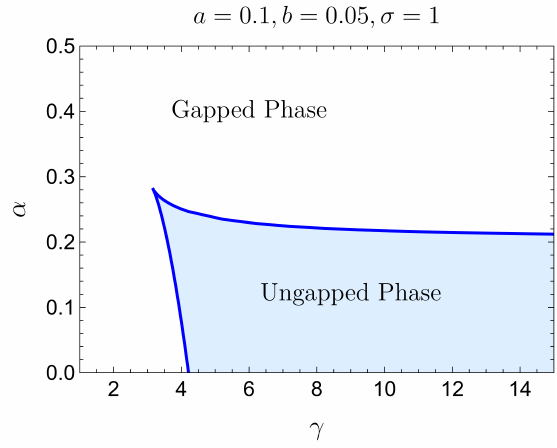} 
    \caption{\small Support of the spectral density (left), where the pole at $z=-1/\alpha$ lies outside $\mathcal{C}$, and the corresponding ungapped--gapped phase boundaries (right) in the small-$\alpha$ regime.}
    \label{fig:Support_Phase_Small_alpha}
\end{figure}
In this phase, the spectral density and $\mathscr{N}$ are determined from the equation of motion \eqref{eq:EOM_ungapped}. We obtain $\mathscr{N} = 0$, with the spectral density given by
\begin{equation}
    \rho(z)=1+a\,z+b\,z^{-1}+\sigma\left(1+\dfrac{1}{1+\gamma\,z}-\dfrac{1}{1+\alpha\,z}\right)
\end{equation}
The support $\mathcal{C}$ of the eigenvalue distribution is obtained from Eq.~\eqref{Contour}.
\begin{equation}
    \alpha |z|^{1+\frac{1}{\sigma}}\,e^{\frac{1}{\sigma}\, \text{Re}\,z \left( a - \frac{b}{|z|^2} \right)} = \Big|z + \dfrac{1}{\alpha}\Big|^{-1} \, \Big|z +\dfrac{1}{\gamma}\Big|
\end{equation}
The $\mathrm{SU}(N)$ constraint Eq.~\eqref{eq:SU} takes the form
\begin{equation}\label{eq:SUN_small_alpha}
\frac{b}{r_0} - a r_0 
+ (1 + \sigma)\ln r_0 
+ \sigma \ln \gamma 
+ \sigma\ln\!\left(\frac{1-r_0 \alpha}{r_0 \gamma - 1}\right)=0
\end{equation}
The above equation is transcendental, so we solve it numerically. 

 We now examine how this phase evolves as $\alpha$ increases. The pole of $\rho(z)$ at $z=-1/\alpha$ is required to remain outside the contour $\mathcal{C}$; however, this requirement fails for sufficiently large $\alpha$. As a result, the spectral density may develop a zero on the negative real axis at $z = -r_0$, signaling the onset of a gap. At this point, the support of the eigenvalue distribution splits, and the system enters the gapped phase. Imposing the condition $\rho(-r_0)=0$ then determines the phase boundary.
\begin{equation}
\alpha=\frac{b\left(-1 + r_0 \gamma\right)+ r_0\left(1 - r_0 \gamma + a r_0 \left(-1 + r_0 \gamma\right) + \sigma\right)}{r_0\left[b\left(-1 + r_0 \gamma\right)+ r_0\left(1 + a r_0 \left(-1 + r_0 \gamma\right) + 2\sigma - r_0 \gamma (1 + \sigma)\right)\right]}
\end{equation}
The winding Wilson loops are given by
\begin{equation}\label{eq:Wilson_in_small_alpha}
W_n := \frac{1}{N}\langle \mathrm{tr}\,U^n\rangle
= \oint_{\mathcal C}\frac{dz}{2\pi i z}\,\rho(z)\,z^n
= 
\begin{cases}
b+\dfrac{\sigma}{\gamma} & n= 1\\[3pt]
(-1)^{n+1}\sigma\gamma^{-n} & n\geq 2 \\[3pt]
a+\sigma\alpha & n = -1 \\[3pt]
(-1)^{n+1}\sigma\alpha^{|n|} & n\leq -2 .
\end{cases}
\end{equation}
 A notable feature of the present model is that
\begin{equation}
W_n \neq W_{-n},
\end{equation}
which directly signals the presence of a sign problem. This asymmetry is expected, since the model's effective potential is complex.

Using the same method as in the $\mu = 0$ case, the free energy in this phase is given by
\begin{equation}
    F = -a\,b - b\,\sigma\alpha - \frac{a\,\sigma}{\gamma}
    + \sigma^2 \ln\!\left(1 - \frac{\alpha}{\gamma}\right).
\end{equation}
\textbf{Remarks:} As discussed in Sec.~\ref{sec:Intro}, the observables of the model reduce to the expected results in the relevant limits, showing that the model consistently reproduces known cases.

Using the relations $a = z_b\, W_{-1}$ and $b = z_b\, W_{1}$ derived in Sec.~\ref{sec:Truncation}, and substituting them into Eq.~\eqref{eq:Wilson_in_small_alpha}, we obtain the self-consistent Wilson loops
\begin{equation}\label{eq:Wilson_mu_0_ungapped}
W_{-1} = \frac{\sigma \alpha}{1 - z_b(\beta/R)}, 
\qquad  
W_{1} = \frac{\sigma \gamma^{-1}}{1 - z_b(\beta/R)}.
\end{equation}
This result is consistent with \cite{Christensen:2012km}.

\subsubsection{Large $\alpha$ ($\mu>\varepsilon$)}
\begin{figure}[t]
    \centering
    \includegraphics[width=0.49\linewidth]{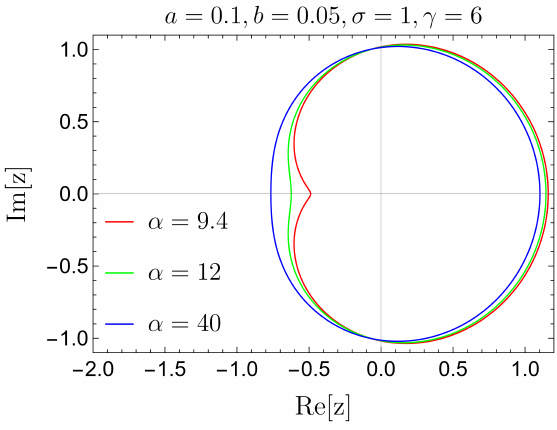}
    \includegraphics[width=0.48\linewidth]{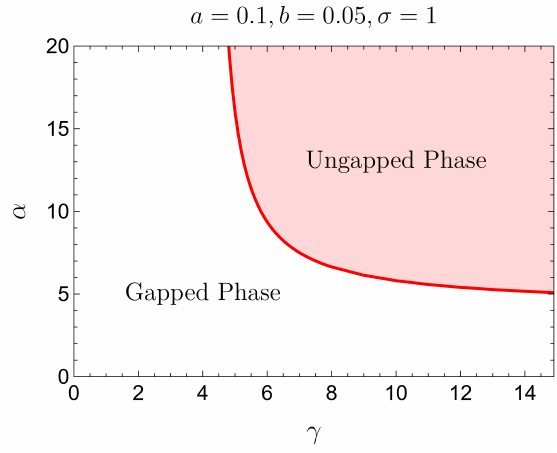} 
    \caption{\small Support of the spectral density (left), where all poles lie inside the contour, and the corresponding ungapped--gapped phase boundaries (right) in the large-$\alpha$ regime.}
    \label{fig:Support_Phase_Small_alpha}
\end{figure}
The large-$\alpha$ phase can be analyzed in a similar manner. In this phase, $\mathscr{N} = \sigma$, and the spectral density takes the form
\begin{equation}
    \rho(z)=1+a\,z+b\,z^{-1}+\sigma\left(\dfrac{1}{1+\gamma\,z}+\dfrac{1}{1+\alpha\,z}\right)
\end{equation}
The support $\mathcal{C}$ of the eigenvalue distribution is
\begin{equation}
    |z|^{2+\frac{1}{\sigma}}\,e^{\frac{1}{\sigma}\, \text{Re}\,z \left( a - \frac{b}{|z|^2} \right)} = \Big|z + \dfrac{1}{\alpha}\Big| \, \Big|z +\dfrac{1}{\gamma}\Big|
\end{equation}
The SU($N$) constraint equation is
\begin{equation}\label{eq:SUN_large_alpha}
\frac{b}{r_0} - a r_0 
+ (1 + 2\sigma)\ln r_0 
+ \sigma \ln(\alpha \gamma) 
- \sigma\ln\!\left[(1-r_0 \alpha) (r_0 \gamma - 1)\right]=0
\end{equation}
The above equation is transcendental, so we solve it numerically.

We now examine the evolution of this phase as $\alpha$ decreases. The poles of $\rho(z)$ at $z=-1/\alpha$ and $z=-1/\gamma$ are required to lie inside the contour $\mathcal{C}$; however, this condition fails for sufficiently small $\alpha$. Thus, the phase boundary in this phase is 
\begin{equation}
\alpha=\frac{
\left(b + r_0(-1 + a r_0)\right)\left(-1 + r_0 \gamma\right) 
+ r_0 (2 - r_0 \gamma)\sigma
}{
r_0 \left[
\left(b + r_0(-1 + a r_0)\right)\left(-1 + r_0 \gamma\right) 
+ r_0 \sigma
\right]
}
\end{equation}
The winding Wilson loops are expressed as
\begin{equation}
W_n = 
\begin{cases}
b+\sigma\left(\dfrac{1}{\gamma}+\dfrac{1}{\alpha}\right) & n= 1\\[3pt]
(-1)^{n+1}\sigma\left(\dfrac{1}{\gamma^n}+\dfrac{1}{\alpha^n}\right) & n\geq 2 \\[3pt]
a & n = -1 \\[3pt]
0 & n\leq -2 .
\end{cases}
\end{equation}
The free energy in this phase is 
\begin{equation}
    F=-a\,b-\sigma\ln{\alpha-a\,\sigma\left(\dfrac{1}{\alpha}+\dfrac{1}{\gamma}\right)}
\end{equation}
\\
\textbf{Remarks:} As discussed in Sec.~\ref{sec:Intro}, the observables of the model reduce to the expected results in the relevant limits, showing that the model consistently reproduces known cases. In particular, in the limit of the GWW model and in the $\mu=0$ limit of QCD, the two ungapped phases of the present model merge into a single ungapped phase.

Using the relations $a = z_b\,W_{-1}$ and $b = z_b\,W_{1}$, the self-consistent values of the Wilson loops are given by
\begin{equation}\label{eq:Wilson mu=0 ungapped}
    W_{-1}=0, \qquad  W_{1}=\frac{\sigma (\alpha+\gamma)}{\alpha\gamma\left(1 - z_b(\beta/R)\right)}.
\end{equation}

\subsection{Gapped Phase}
 Using Eq.~\eqref{eq:g(z)}, the function $g(z)$ is obtained as
 \begin{equation}
     g(z)=a+\dfrac{b}{|\mathfrak{z}|\,z}+\sigma\left(\frac{1}{A_\gamma(1+\gamma z)}+\frac{1}{A_\alpha(1+\alpha z)}\right)
 \end{equation}
where $|\mathfrak{z}|$ denotes the magnitude of the endpoint $\mathfrak{z}$. We have introduced the shorthand notations
\begin{equation*}
  A_\alpha \equiv \left|\frac{1}{\alpha}+\mathfrak{z}\right|, 
  \qquad 
  A_\gamma \equiv \left|\frac{1}{\gamma}+\mathfrak{z}\right|.
\end{equation*}
To determine the remaining parameters $\mathfrak{z}$, $\mathfrak{z}^\ast$ and $\mathscr{N}$, we require three independent conditions. 
Since $g(z)$ is now fixed, two of them follow from the asymptotic behaviour of the resolvent \eqref{endpoint}, yielding
\begin{figure}[t]
    \centering
    \includegraphics[width=0.5\linewidth]{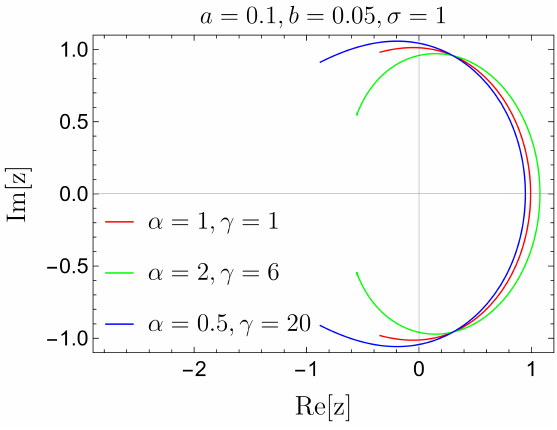}
    \caption{\small Support of the spectral function $\rho(z)$ in the gapped phase.}
    \label{fig:gappedContour}
\end{figure}
\begin{equation}\label{eq:endpoint_Constraints_finite_mu}
    \begin{split}    
    \mathscr{N}-|\mathfrak{z}|\left(a-\dfrac{b\,\text{Re}\,\mathfrak{z}}{|\mathfrak{z}|^3}\right)+\sigma-\sigma |\mathfrak{z}| \!\left(\frac{1}{A_\alpha}+\frac{1}{A_\gamma}\right) = -1, 
    \\
    \mathscr{N}-a\,\text{Re}\, \mathfrak{z}+\dfrac{b}{|\mathfrak{z}|}-\sigma+\sigma\!\left(\frac{1}{\alpha A_\alpha}+\frac{1}{\gamma A_\gamma}\right) = 1 .
    \end{split}
\end{equation}
The third condition is provided by the SU$(N)$ constraint $\langle \ln U \rangle = 0$. Using \eqref{eq:openToClosed}, it can be written as
\begin{equation}
    \oint_{\mathcal C^\ast} \frac{dz}{4\pi i z}\, R(z)\, \ln z = 0 .
\end{equation}
Deforming the contour $\mathcal C^\ast$ as in Fig.~\ref{fig:deformed contour}, the SU$(N)$ constraint becomes
\begin{equation}
    \left(\oint_{0}+\oint_{-\frac{1}{\alpha}}+\oint_{-\frac{1}{\gamma}}-\oint_{\infty}\right)
    \frac{dz}{2\pi i z}\, R(z)\ln z
    + \int_{\epsilon\to 0}^{\Gamma\to \infty}\frac{dx}{x}\, R(-x) = 0 .
\end{equation}
The final form of the SU$(N)$ constraint is therefore
\begin{equation}\begin{aligned}\label{eq:SUN_gapped_finite_mu}
    & \left(\dfrac{a}{|\mathfrak{z}|}-b\,|\mathfrak{z}|\right)\left[\left(1-\dfrac{\text{Re}\,\mathfrak{z}}{|\mathfrak{z}|}\right)+\left(1+\dfrac{\text{Re}\,\mathfrak{z}}{|\mathfrak{z}|}\right) \ln \cos^2 \frac{\phi}{2}\right]+\left(\dfrac{a}{|\mathfrak{z}|}+b\,|\mathfrak{z}|\right)\left(1-\dfrac{\text{Re}\,\mathfrak{z}}{|\mathfrak{z}|}\right)\ln{|\mathfrak{z}|}\\
    & +\sigma\left(\dfrac{1}{A_\alpha} \left[ \left( |\mathfrak{z}| + \dfrac{1}{\alpha} \right) \ln |\mathfrak{z}| - \left( |\mathfrak{z}| - \dfrac{1}{\alpha} \right) \ln \cos^2 \frac{\phi}{2} \right] -  \text{arcsinh} \dfrac{(\alpha - |\mathfrak{z}|^{-1}) A_\alpha}{1 + \cos \phi} + \ln \alpha\right)\\
    & + \sigma\left(\dfrac{1}{A_\gamma} \left[ \left( |\mathfrak{z}| + \dfrac{1}{\gamma} \right) \ln |\mathfrak{z}| - \left( |\mathfrak{z}| - \dfrac{1}{\gamma} \right) \ln \cos^2 \frac{\phi}{2} \right] -  \text{arcsinh} \dfrac{(\gamma - |\mathfrak{z}|^{-1}) A_\gamma}{1 + \cos \phi} + \ln \gamma\right) = 0,
\end{aligned}\end{equation}
where $\phi \in (0,\pi)$ is the argument of the endpoint $\mathfrak{z}=|\mathfrak{z}|e^{i\phi}$. 
The system consisting of \eqref{eq:endpoint_Constraints_finite_mu} together with the SU$(N)$ constraint determines $\mathscr{N}$ as well as the endpoints $\mathfrak{z}$ and $\mathfrak{z}^\ast$.

Since the SU$(N)$ constraint is transcendental, observables in this phase do not admit closed-form expressions in terms of the parameters $a,\,b,\,\sigma,\,\alpha,\,\gamma$. In particular, the single-winding Wilson loop expectation values
\begin{equation}
\begin{split}
    W_1 =\dfrac{a\,(\text{Im}\,\mathfrak{z})^2}{4}+\dfrac{b}{2}\left(1-\dfrac{\text{Re}\,\mathfrak{z}}{|\mathfrak{z}|}\right)
   +\dfrac{\sigma}{2}\left(\dfrac{1}{\alpha }+\dfrac{1}{\gamma }\right)-\dfrac{\sigma}{2 |\mathfrak{z}\, \alpha + 1|} \left(\dfrac{1}{\alpha} + \text{Re}\, \mathfrak{z} \right)\\
   - \dfrac{\sigma}{2 |\mathfrak{z} \gamma  + 1|} \left(\dfrac{1}{\gamma} + \text{Re}\, \mathfrak{z} \right)
\end{split}
\end{equation}
\begin{equation}
\begin{split}
    W_{-1} = \dfrac{b\,(\text{Im}\,\mathfrak{z})^2}{4\,|\mathfrak{z}|^4}+\dfrac{a}{2}\left(1-\dfrac{\text{Re}\,\mathfrak{z}}{|\mathfrak{z}|}\right)+\dfrac{\sigma}{2}(\alpha +\gamma )-\dfrac{\sigma \alpha }{2 |\alpha + 1/\mathfrak{z}|} \left( \alpha  + \dfrac{\text{Re}\, \mathfrak{z}}{|\mathfrak{z}|^2} \right)\\
    - \dfrac{\sigma \gamma }{2 |\gamma + 1/\mathfrak{z}|} \left( \gamma  + \dfrac{\text{Re}\, \mathfrak{z}}{|\mathfrak{z}|^2} \right)
\end{split}
\end{equation}
depend on the endpoints $\mathfrak{z}$ and $\mathfrak{z}^\ast$, which themselves depend transcendentally on the model parameters.
\\
\\
\textbf{Remarks:} We now discuss the limiting case of the model in the gapped phase. In this regime, the corresponding observables are consistently recovered upon taking appropriate limits, as discussed in Sec. \ref{sec:Intro}. For example, when $\sigma=0$ and $a=b$, the model reduces to the GWW model. In this limit, the SU$(N)$ constraint equation \eqref{eq:SUN_gapped_finite_mu} implies $|\mathfrak{z}|=1$, so that the eigenvalues are distributed along the unit circle. Moreover, using the endpoint constraints \eqref{eq:endpoint_Constraints_finite_mu}, we find that $\mathscr{N}=0$ and
\begin{equation}
\phi=2\sin^{-1}\!\left(\frac{1}{2\lambda}\right)^{1/2}.
\end{equation}
The spectral density and all physical observables reduce consistently to their GWW counterparts. As an illustration, consider the Wilson loop,
\begin{equation}
W_{\pm1}=\frac{\lambda}{2}\left(1-\mathrm{Re}\,\mathfrak{z}+\dfrac{(\mathrm{Im}\,\mathfrak{z})^2}{2}\right),
\qquad 
\mathrm{Re}\,\mathfrak{z}=\cos{\phi},\quad \mathrm{Im}\,\mathfrak{z}=\sin{\phi}.
\end{equation}
Substituting the above expression for $\phi$, we obtain
\begin{equation}
W_{\pm1}=1-\frac{1}{4\lambda},
\end{equation}
in exact agreement with the GWW result. We similarly recover the expected observables in the other limiting cases.

All plots in the ungapped and gapped phases were generated using \textsc{Mathematica}. We discuss the corresponding phase transitions in the next section.

\subsection{Phase transitions}
\begin{figure}[t]
    \centering
    \includegraphics[width=.48\textwidth]{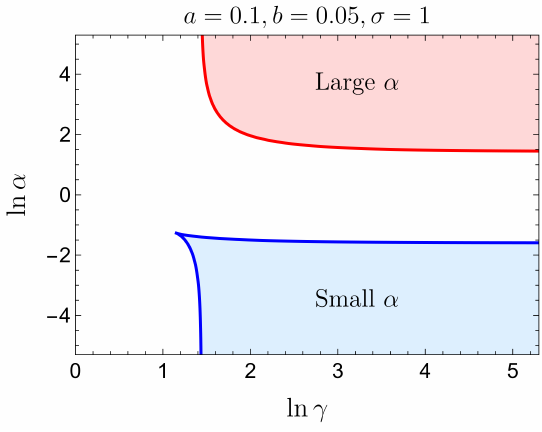}
    \,
    \includegraphics[width=.48\textwidth]{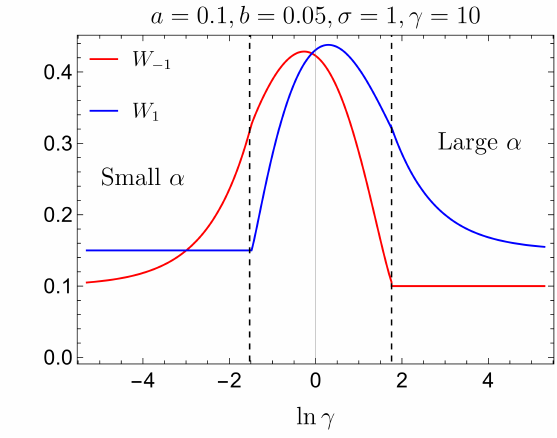}
    \caption{\small The left panel shows the phase diagram, where the colored region shows the ungapped phases, while the white region shows the gapped phase, and the right panel illustrates how the Wilson loop behaves across the transition. }
    \label{fig:Combine Phase and Wilson}
\end{figure}
The full phase diagram and the corresponding Wilson loop expectation values are shown in Fig.~\ref{fig:Combine Phase and Wilson}, where the left panel displays the phase diagram and the right panel illustrates the Wilson loop behavior across the transition. Interestingly, there are no direct transitions between ungapped phases; instead, any transition between ungapped phases must proceed through an intermediate gapped phase.

To determine the order of the phase transition, one must analyze the behavior at the critical point. The transition occurs when
\begin{equation}
\arg(\mathfrak{z}) \to \pi, \qquad |\mathfrak{z}| \to r_{0},
\end{equation}
where $r_{0}$ is the radius on the negative real axis at which the gap opens.

Using the $SU(N)$ constraints in Eqs.~\eqref{eq:SUN_small_alpha} and \eqref{eq:SUN_large_alpha}, one determines the value of $|\mathfrak{z}|$ at the transition points numerically in both the small-$\alpha$ and large-$\alpha$ ungapped phases. Subsequently, employing the endpoint condition in Eq.~\eqref{eq:endpoint_Constraints_finite_mu} for the gapped phase, one finds that as the system approaches the transition from the gapped phase to the small-$\alpha$ ungapped phase, $\mathscr{N} \to 0$, whereas approaching the large-$\alpha$ ungapped phase yields $\mathscr{N} \to \sigma$.

Since $\mathscr{N}$ remains continuous across the transition points, the transition is not first order, indicating instead a continuous phase transition of at least second order \cite{hands_qcd_2010}. Similarly, one finds that the Wilson loop is also continuous across the transition.

\section{Conclusion and Discussion}
In this work, we examine the large-$N$ limit of a QCD-inspired unitary matrix model. The analysis is carried out in two parts: the $\mu = 0$ case, where the potential becomes real, and the finite-$\mu$ case, where the potential is complex. In both cases, the ungapped (confined) phase successfully reproduces the low-temperature features of QCD. 

The gapped (deconfined) phase, however, is considerably more involved and does not admit a complete analytic solution; consequently, the corresponding QCD-like features cannot be fully reproduced in this phase. Nevertheless, the model provides a more general framework, as discussed in Sec.~\ref{sec:Intro}. For $\mu = 0$, the model exhibits a $3^{rd}$ order phase transition, while for finite~$\mu$, it shows a continuous phase transition of at least second order.

Analytic expressions for all thermodynamic observables are obtained in the ungapped phases. In contrast, in the gapped phase, the Wilson loop value can be determined only partially analytically and otherwise requires numerical evaluation. These results provide further insight into the phase structure of non-Hermitian extensions of unitary matrix models and may offer useful analogies for understanding complex actions in QCD-like theories. Possible extensions of this work include incorporating finite-$N$ effects and exploring the behavior under more general complex deformations of the potential.
\newpage
\bibliography{Citations}

\end{document}